# ROTATIONAL SPECTRA OF THE EXOTIC AND NON-EXOTIC BARYONS

David Akers*
*Lockheed Martin Corporation, Dept. 6F2P, Bldg. 660, Mail Zone 6620,*
1011 Lockheed Way, Palmdale, CA 93599
*Email address: David.Akers@lmco.com

**Abstract**

In the present paper, the rotational states of the baryons, dibaryons and baryon octets are shown. A comparison of the S = +1 theta baryons is made to the spectrum of the narrow width dibaryons. From this comparison, the rotational spectra of the exotic theta baryons are proposed. We analyze the importance of the 70 MeV quantum proposed by Mac Gregor and its relationship in the location of exotic baryons belonging to the antidecuplet. Antidecuplet mass separations are suggested on the order of the pion mass or ≈ 120 – 150 MeV from symmetry to the spectrums of the baryon decuplet and charmed baryons. A mass separation is possible on the order of the muon mass or ≈ 108 MeV as recently suggested by Diakonov and Petrov. However, the mass separation of 108 MeV introduces an asymmetry with respect to the mass separations as found in the baryon decuplet.

## INTRODUCTION

A study is presented of the rotational spectra of exotic and non-exotic baryons. From a previous analysis of the experimental data [1-4], it was shown that the rotational energies of the baryons follow the L(L + 1) interval rule and that the average rotational energy of the baryons is approximately 30 MeV. This analysis is based upon the existence of a 70 MeV quantum, or multiples thereof, as a fundamental vibration or excitation from the baryon ground states [3]. This 70 MeV quantum has been derived from a modified QCD Lagrangian [5] and shown to fit the Nambu empirical mass formula [6]. The Nambu formula consists of quanta 35, 70, 105, 140 MeV, … for particle masses. Mac Gregor has extensively studied the 70 MeV quantum, or multiples thereof, as a fundamental vibration or excitation. His notation included: m = 70 MeV, B = 140 MeV or the mass of the pion, F = 210 MeV or twice the muon mass, and X = 420 MeV [7]. Mac Gregor proposed a rotational model of the elementary particles based upon a 70 MeV grid [3]. We shall utilize evidence for the existence of the 70 MeV quantum in the present work.

We continue with our analysis of the rotational states of the baryons from a previous study [4]. In the present paper, the rotational states of the baryons, dibaryons and baryon octets are shown. A comparison of the S = +1 theta baryons is made to the spectrum of the narrow width dibaryons. From this comparison, the rotational spectra of the exotic theta baryons are proposed. We analyze the importance of the 70 MeV quantum proposed by Mac Gregor and its relationship in the location of exotic baryons belonging to the antidecuplet. Antidecuplet mass separations are suggested on the order of the pion mass or ≈ 120 – 150 MeV from symmetry to the spectrums of the baryon decuplet and charmed baryons. A mass separation is possible on the order of the muon mass or ≈ 108 MeV as recently suggested by Diakonov and Petrov [8]. However, the 108 MeV spacing suggests an asymmetry with respect to the mass separations as found in the baryon decuplet.

## BARYON ROTATIONAL STATES

Mac Gregor has demonstrated how to accurately bridge the gap between hadronic and nuclear domains [2]. By an application of nonadiabatic rotational bands, Mac Gregor determined a proper understanding of particle resonances. The nature of the nonadiabatic rotations starts from the rotational Hamiltonian [2]:

$$H = (\hbar)^2 \mathbf{L}^2/2I \, , \tag{1}$$

where **L** is the angular momentum operator and $I$ is the moment of inertia. The angular momentum operator can be written as

$$\mathbf{L} = \mathbf{J} - \mathbf{S}, \tag{2}$$

where **J** and **S** are the total angular momentum and spin angular momentum operators. The angular momentum operator in Eq. (2) can be written as

$$\mathbf{L}^2 = (\mathbf{J} - \mathbf{S})^2 = \mathbf{J}^2 + \mathbf{S}^2 - 2\mathbf{J} \cdot \mathbf{S} \, . \tag{3}$$

Mac Gregor noted that the Coriolis term $\mathbf{J} \cdot \mathbf{S}$ tended to vanish if the rotating system satisfied a few conditions [2]. One of these conditions is the intrinsic spin S does not have the value ½. Mesons would fit into this category. If the Coriolis term vanishes, then the rotational energies of the mesons would follow a J(J + 1) interval rule:

$$E\,(J) = E_0 + (\hbar)^2 \, J(J + 1)/2I \, ,$$

$$E_{rot} = (\hbar)^2 /2I \, , \tag{4}$$

where $E_0$ is the bandhead energy. The bandhead energy is the ground state of the particular meson series. On the other hand, when the Coriolis term does not vanish for rotational levels, the particle resonances follow the L(L + 1) interval rule and the rotational energies are

$$E\ (J) = E_0 + (\hbar)^2\ L(L + 1)/2I\ ,$$

$$E_{rot} = (\hbar)^2\ /2I\ . \qquad (5)$$

The baryons therefore follow the L(L + 1) interval rule of Eq. (5). Likewise, Mac Gregor showed Eq. (5) is the correct equation for all light nuclei with $A \leq 20$ [2].

The rotational spectra of the baryons have been previously tabulated [4]. The known and not-so-well established baryons were sorted by angular momentum L and on an energy scale with a 70 MeV grid. From the tabulated data [4], we plot the rotational states of the nucleons in Fig. 1. In Fig. 1, we note the nucleon ground state as N(939). The excitation levels directly above the N(939) nucleon are separated from the ground states by 420 MeV, 596 MeV and 711 MeV, respectively for N(1359)S, N(1535)S and N(1650)S. The N(1359)S is predicted to exist, for which there is some evidence [9]. The excitation energy 420 MeV is a multiple of 70 MeV. The N(1535)S is approximately 595 MeV above the N(939), which represents 595 = 8(70) + 35 MeV. The N(1650)S is about 700 MeV above the nucleon N(939) or 700 = 10(70) MeV. The patterns of energy separation from the Nambu formula are not accidental [10]. These are essentially excitations on the order of the muon and pion masses. Therefore, many baryons will have decay channels, involving muons and pions as by-products. The nucleons decay predominately with pions as by-products.

Although we can understand the rotational states of the nucleons based upon Eq.(5), there has been recent evidence for even lighter nucleons [11, 12]. There is evidence for N(966)S and N(987)P from Fil'kov *et al.* [11]. The N(966)S is about 27 MeV above the N(939). This energy separation is on the order of 35 MeV, meaning that a 35 MeV excitation from N(939) would result in 8 MeV or 0.8 % binding energy for the N(966). On the other hand, the N(987)P is a rotational state with 47 MeV in energy above the N(939). From Eq.(5), this energy separation would represent a rotational energy of $E_{rot}$ = 23.5 MeV for N(987)P near the bandhead energy of 939 MeV. This rotational energy is rather low for the average rotational energy of 30 MeV for the baryons [4]. However, we are dealing with a rather low vibration or excitation quantum of 35 MeV.

In addition, the Saclay group has recently reported evidence for N(1004), N(1044), and N(1094) [12]. There is no reported information on spins and parities for these nucleons. Therefore, we can estimate the possible rotational spectra of these light nucleons from our knowledge of an excitation m = 70 MeV quantum from the N(939) ground state. Our analysis is shown in Fig. 2. In Fig. 2, we show the rotational spectra of the light nucleons along with a suggested spectrum for the N(1004), N(1044), and N(1094). For a S-state, the N(1004) would represent an approximate 70 MeV excitation above the N(939). The dashed red line represents this excitation level in Fig. 2. The 140 MeV excitation level is also represented by a dashed red line above the N(939). However, if N(1004) is a P-state, then this would represent a rotational energy of $E_{rot}$ = 32.5 MeV. This is about the average baryon rotational energy [4]. The N(1044) is about 105 MeV or a muon mass excitation above the N(939). If N(1044) is the P-state associated with a N(1004), then the rotational energy of N(1044)P would be 20 MeV, which is a typical dibaryon rotational energy [13]. If N(1094) is a P-state associated with N(1004)S, then the rotational energy would be 42.5 MeV, which is typical for the rotational energies of the mesons [2-4]. This is a rather high value for the nucleons, and therefore we suggest that N(1094) is a S-state as shown in Fig. 2. The N(1149)S would be the next predicted nucleon in Fig. 2 from our analysis based upon Eq.(5).

The rotational energy spectra of all the baryons have been tabulated in Ref. [4]. From these tabulated data, we can plot the rotational states of all the baryons. These are shown in Figs. 3 to 7. In Fig. 3, we have plotted the rotational states of the baryon octets. In Figs. 4 to 7, we have plotted the rotational states of the delta, lambda, sigma and cascade baryons, respectively. The solid lines represent baryons taken from the *Review of Particle Properties (2002)* [14]. The dashed lines represent predicted states based upon Eq.(5) for each baryon series. The bandhead states of energy $E_0$ are the S-states as shown in these figures. From these bandhead states, the energy states have been calculated and compared to the experimental data [4]. The experimental data for these rotational spectra are found to be linear for L(L + 1). The results of our analysis are in agreement with the earlier findings of Mac Gregor [1-3].

## EXOTIC BARYONS

In 2003, several collaborations have discovered an exotic baryon with five quarks [15-18]. The recently discovered exotic baryon has been named $\theta^+(1540)$. This baryon was a seminal prediction by Diakonov, Petrov, and Polyakov [19]. These authors utilized a Skyrme-inspired quark soliton model. However, the idea that baryons can be constructed from the combinations of three quark states (qqq), five quark states (qqqq-qbar), etc. originated with Gell-Mann [20].

In the paper by Diakonov, Petrov and Polyakov [19], there are predictions of particle masses for the antidecuplet. These authors originally associated a $\theta^+(1530)$ with N(1710), $\Sigma(1890)$, and $\Xi(2070)$. In a most recent work [21], Diakonov and Petrov suggest an equal mass splitting of 108 MeV for the antidecuplet. These authors now predict a new nucleon at 1650-1690 MeV and a new $\Sigma$ baryon at 1760-1810 MeV, belonging to the antidecuplet with $\theta^+(1540)$. In addition, they also place a recently discovered $\Xi_{3/2}(1862)$ as a member in the *same* antidecuplet. It remains to be seen from experimental efforts which nucleon is associated with the new S = + 1 baryon $\theta^+(1540)$. However, if these authors are correct, then the 108 MeV spacing would represent an asymmetry with respect to the known mass splitting of the decuplet, which is on the order of the pion mass:

$$M_{\Omega} - M_{\Xi*} = 142 \text{ MeV};$$

$$M_{\Xi*} - M_{\Sigma*} = 145 \text{ MeV, and} \qquad (6)$$

$$M_{\Sigma*} - M_{\Delta} = 153 \text{ MeV}.$$

Thus, the SU(3) decuplet has an "approximate" equal spacing rule as shown in Eq.(6).

We postulate that the antidecuplet has an equal spacing rule on the order of the pion mass in comparison to the rule for the decuplet. This reasoning is based upon symmetry with respect to what we find in the mass separations for the baryon decuplet and for the charmed baryons. In addition, we will show that the ratios for the decay widths of the antidecuplet members are in general agreement with the ratio for the decuplet members.

We start with a study of the rotational states of the p-p dibaryons from the work of Mac Gregor [13]. In Fig. 8, we show the rotational levels of the *narrow width* p-p dibaryons. This figure is generated from Mac Gregor's Fig. 1 in Ref. [13]. The rotational energy of the p-p dibaryons is 20 MeV. The first excitation state, (2020)S, is 143 MeV or a pion mass above the p-p(1877) ground state as shown in Fig. 8. The dibaryons are known to follow the rule of 70, 140, and 210 MeV in excitation energy from the ground states, and this is shown in Fig. 9.

In Fig. 9, we indicate the strangeness of the dibaryons at the bottom of the figure. This figure is generated from Mac Gregor's Fig. 8 in Ref. [10]. We show the $K^+N(1433)$ ground state next to the dibaryons for comparison. We note the excitation energies of 70, 140, and 210 MeV for the dibaryons in Fig. 9. From symmetry, we can expect the vibration or excitation states of the theta baryons to follow a similar pattern. This pattern is shown as dashed lines in Fig. 9. Therefore, the S-state excitations are 1503, 1573, and 1643 MeV. From the fact that the rotational energies of the dibaryons are 20 MeV [13], we postulate that the theta baryons will also have a rotational energy $E_{rot} \approx 20$ MeV. With the S-state excitations of Fig. 9, we build a rotational spectrum for the theta baryons, having a rotational energy $E_{rot} \approx 20$ MeV, as shown in Fig. 10. If experiments show that the $\theta^+(1540)$ has a spin-parity $J^P = (1/2)^-$, then it is a S-state at 107 MeV above the ground state $K^+N$ or $\theta$ (1433). However, we believe that $\theta^+(1540)$ is most likely a spin-parity $J^P = (1/2)^+$ and is a P-state. The next D-state from $\theta^+(1540)$ is $\theta$(1614)D from the rotational energy of 20 MeV and based upon calculations from Eq.(5). We also postulate the existence of a $\theta$ (1473)P above the threshold $K^+N$(1433). Other postulated states are shown in Fig. 10, and these are based upon Eq.(5) with $E_{rot} \approx 20$ MeV for each rotational series. Our analysis also suggests that there should be even higher rotational states than those shown in Fig. 10, and there is some earlier evidence to support this fact [22-25].

From the spectrum of Fig. 10, we can build a mapping of the theta baryon P-states onto the antidecuplet mass separations. This mapping is shown in Fig. 11. In Fig. 11, the horizontal spacing is on the energy scale of 70 MeV. The vertical mass separations are the order ≈ 120 – 150 MeV from symmetry to the spectrums of the baryon decuplet and charmed baryons. For the charmed baryons, the mass separations are shown in Fig. 12 and are indicates by the arrows. We note the 119, 123.5, and 125 MeV separations for the charmed baryons in Fig. 12. From Eq.(6) and Fig. 12, we can expect to find a similar pattern of mass separations for the antidecuplet members based upon symmetry to the baryon decuplet.

In Fig. 11, we have grouped the antidecuplet based upon $\approx 120 - 150$ MeV separations between members in the vertical. In addition, a $\Xi_{3/2}(1862)$ has been recently discovered [26]. If $\Xi_{3/2}(1862)$ has a spin-parity $J^P = (1/2)^+$, then we can locate this member on the antidecuplet chart of Fig. 11. Based upon the suggested mass separation $\approx 120 - 150$ MeV, we have located $\Xi_{3/2}(1862)$ on a *separate* antidecuplet member line from that for θ(1540). Of course, if Diakonov and Petrov are correct, then the blue triangle will be in alignment with the red square in Fig. 11.

In Fig. 11, the black solid points are experimental data taken from the *Review of Particle Properties (2002)* [14]. The open circles are predicted states for members of each antidecuplet line. The blue solid points have some experimental evidence (see Appendix H in Ref. [3]). The threshold for $K^+N(1433)$ is shown as vertical arrows in Fig. 11. The P-states are possible below this threshold. For example, the θ(1540) may decay with a $\pi^- \pi^+$ channel to θ(1260). We suggest it is *more* likely that there are excitation states above the $K^+N(1433)$ threshold. One of these θ(1473-1493) is shown in Fig. 11. We postulate that θ(1473-1493) and $\Xi_{3/2}(1862)$ are members of the *same* antidecuplet.

Why are θ(1473-1493) and $\Xi_{3/2}(1862)$ members of the *same* antidecuplet? We compare the total widths of the baryon decuplet members Σ(1385)P with Γ35.8 and Ξ(1530)P with Γ9.9 as a ratio:

$$\Gamma\,[\Sigma(1385)]/\,\Gamma\,[\Xi(1530)] = 35.8/9.9 = 3.6 \qquad (7)$$

If the $\Xi_{3/2}(1862)$ of narrow width < 18 MeV is member with the $\Sigma(1738\text{-}1770)$, which has a $\Gamma 80 \pm 30$ MeV, then we have a ratio:

$$\Gamma\,[\Sigma(1738)]/\,\Gamma\,[\Xi(1862)] = 80/18 = 4.4 \qquad (8)$$

The result of Eq.(8) is therefore an upper bound. Comparison of Eq.(8) with Eq.(9) indicates reasonable agreement within bounds. The lower mass 1738 for $\Sigma(1738\text{-}1770)$ suggests a mass separation of 124 MeV with respect to $\Xi_{3/2}(1862)$. Therefore, the nucleon associated with this member antidecuplet line should be N(1610-1633). From the grid of 70 MeV horizontal spacings, we predict that $\theta(1540)$ should be associated with a new N(1655-1680) and a new $\Sigma(1805\text{-}1825)$. In addition, we postulate that a new $\Xi_{3/2}(1930\text{-}1960)$ will be associated with this member line, which is inclusive of $\theta(1540)$ as shown in Fig. 11.

## CONCLUSION

We presented a study of the rotational spectra of exotic and non-exotic baryons. From a previous analysis of the experimental data, it was shown that the rotational energies of the baryons follow the L(L + 1) interval rule and that the average rotational energy of the baryons was approximately 30 MeV. This analysis was based upon the existence of a 70 MeV quantum, or multiples thereof, as a fundamental vibration or excitation from the ground states. A comparison of the S = +1 theta baryons was made to the spectrum of the *narrow width* dibaryons. From this comparison, the rotational spectra of the exotic theta baryons were proposed. We analyzed the importance of the 70 MeV quantum proposed by Mac Gregor and its relationship in the location of exotic baryons belonging to the antidecuplet. Antidecuplet mass separations were suggested on the order of the pion mass or approximately 120 – 150 MeV from symmetry to the spectrums of the baryon decuplet and charmed baryons. A mass separation is possible on the order of the muon mass or approximately 108 MeV as recently suggested by Diakonov and Petrov. However, this suggests an asymmetry with respect to the mass separations as found in the baryon decuplet. From the grid of 70 MeV horizontal spacings, we predicted that θ(1540) should be associated with a new N(1655-1680) and a new Σ(1805-1825). In addition, we postulated that a new $\Xi_{3/2}$(1930-1960) will be associated with this member line, which is inclusive of θ(1540).

*Note added in proof*: Recently, Aslanyan *et al.* discovered evidence of S = + 1 narrow resonances at 1545.1, 1612.5, and 1821.0 MeV in the $pK^0_S$ invariant mass spectrum [27]. The evidence of a 1612 MeV state appears to support our prediction of the θ(1610)P in Fig. 10.

## ACKNOWLEDGEMENT

The author wishes to thank Dr. Malcolm Mac Gregor, retired from the University of California's Lawrence Livermore National Laboratory, for his encouragement to pursue the CQ Model, and he wishes to thank Dr. Paolo Palazzi of CERN for his interest in the work and for e-mail correspondence.

## FIGURE CAPTIONS

Fig. 1. The rotational states of the nucleons are taken from the tabulations in Ref. [4].

Fig. 2. The rotational states of the light nucleons are based upon the calculations from Eq.(5).

Fig. 3. The rotational states of the baryon octets are taken from the tabulations in Ref. [4].

Fig. 4. The rotational spectra of the delta baryons are shown. These states are taken from the tabulations in Ref. [4].

Fig. 5. The rotational states of the lambda baryons are shown. These states are taken from the data tabulated in Ref. [4].

Fig. 6. The rotational states of the sigma baryons are shown. These states are also derived from the data tabulated in Ref. [4].

Fig. 7. The rotational states of the cascade baryons are shown. The states are derived from the tabulations in Ref. [4].

Fig. 8. The rotational levels of the *narrow width* p-p dibaryons are shown. The levels are derived from the work of Mac Gregor [13].

Fig. 9. A comparison is made of the theta baryons to the spectrum of dibaryons. The values of strangeness S are indicated at the bottom of the figure. Note the pattern of 70 MeV excitations in the figure and the symmetry with respect to strangeness S.

Fig. 10. A rotational spectrum is suggested for the theta baryons based upon calculations from Eq.(5) and a rotational energy of 20 MeV.

Fig. 11. A mapping of the rotational spectra of the theta baryons is shown. The chart shows a grid of 70 MeV in the horizontal spacing and a mass separation in the vertical on the order of the pion mass or $\approx 120 - 150$ MeV .

Fig. 12. The spectrum of charmed baryons is shown, revealing the indicated mass separations.

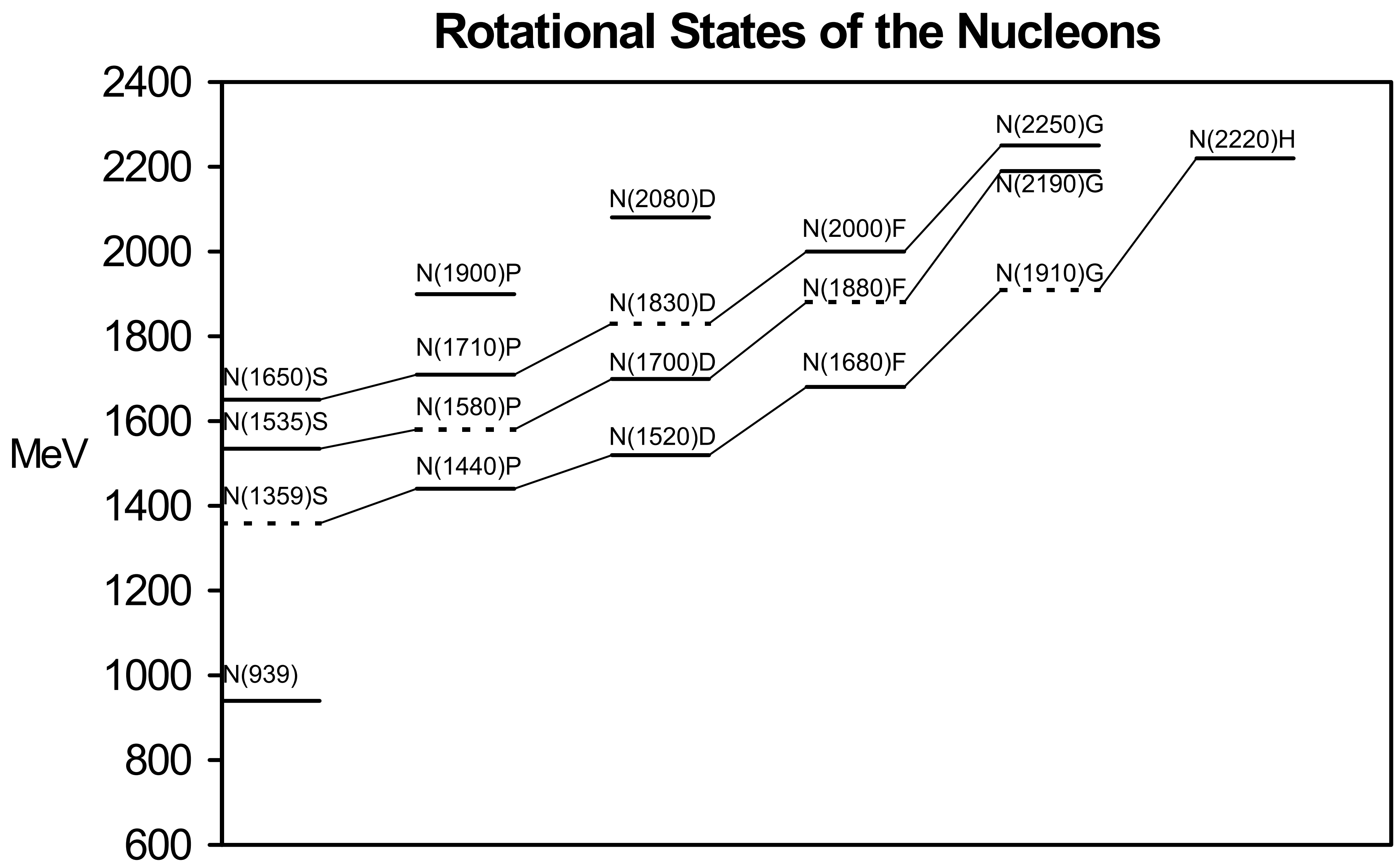

Fig. 1

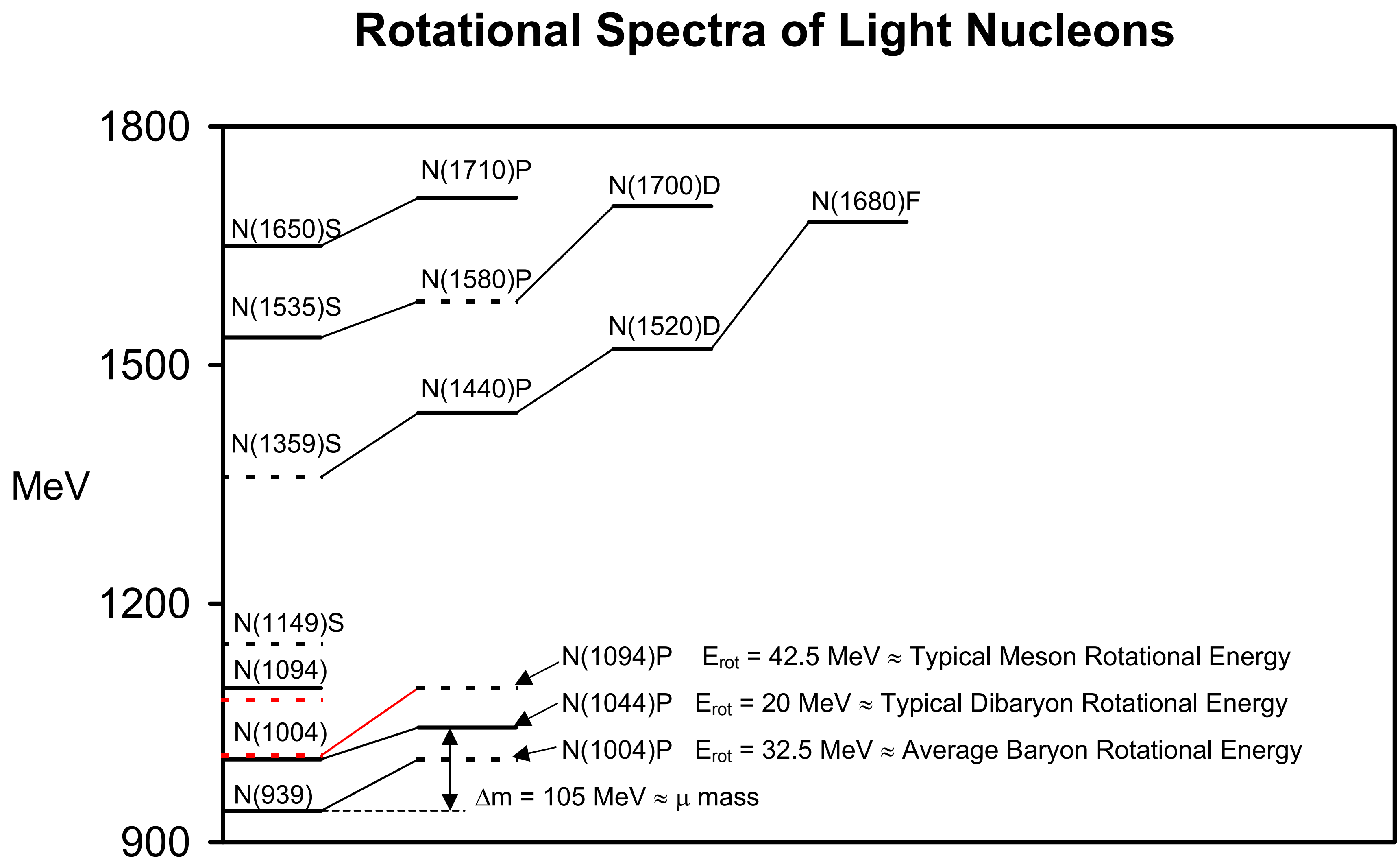
Rotational Spectra of Light Nucleons
1800
1500
1200
900
MeV
N(1710)P
N(1700)D
N(1680)F
N(1650)S
N(1580)P
N(1535)S
N(1520)D
N(1440)P
N(1359)S
N(1149)S
N(1094)
N(1004)
N(939)
N(1094)P   E_rot = 42.5 MeV ≈ Typical Meson Rotational Energy
N(1044)P   E_rot = 20 MeV ≈ Typical Dibaryon Rotational Energy
N(1004)P   E_rot = 32.5 MeV ≈ Average Baryon Rotational Energy
Δm = 105 MeV ≈ μ mass

Fig. 2

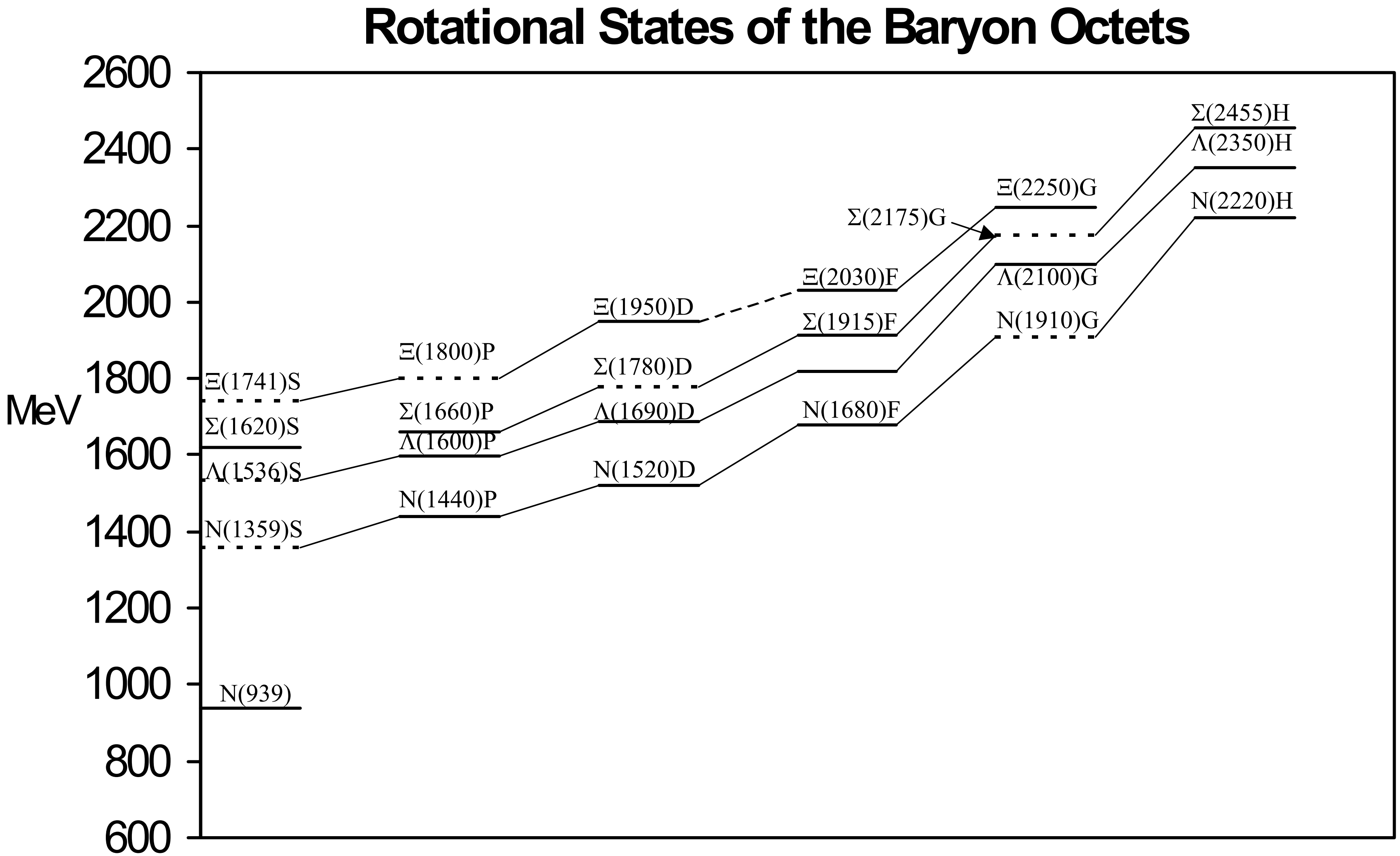

Fig. 3

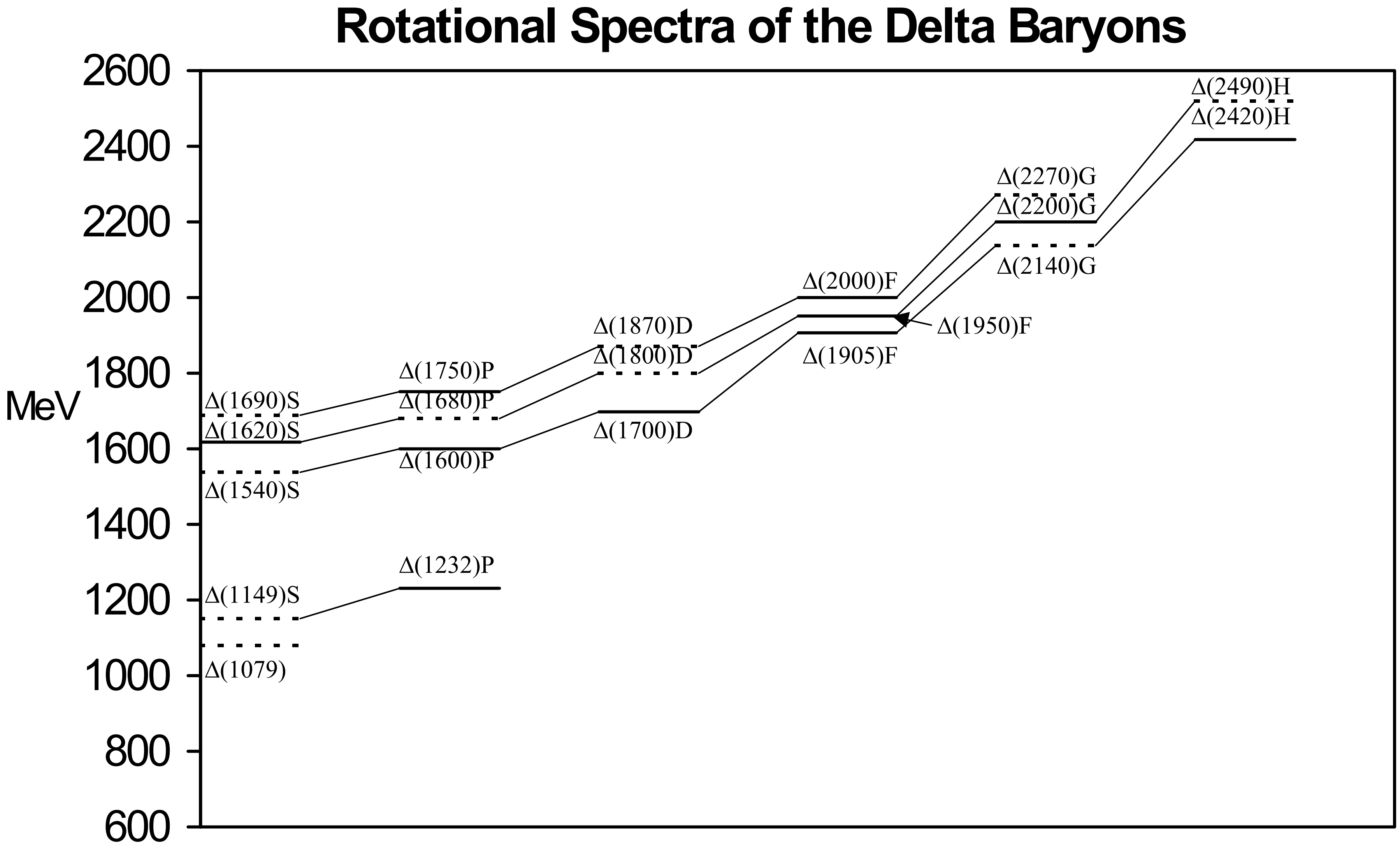

Fig. 4

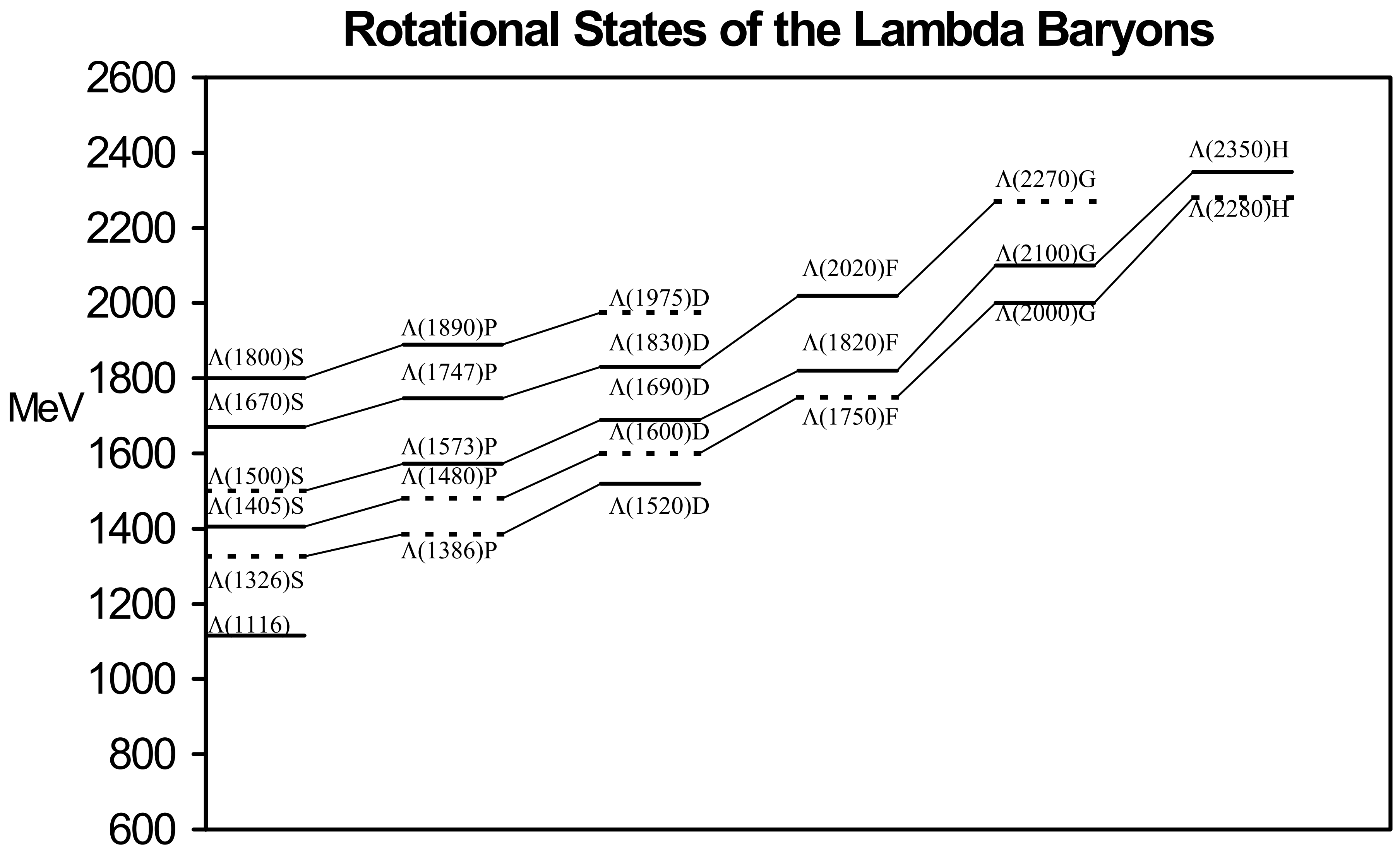

Fig. 5

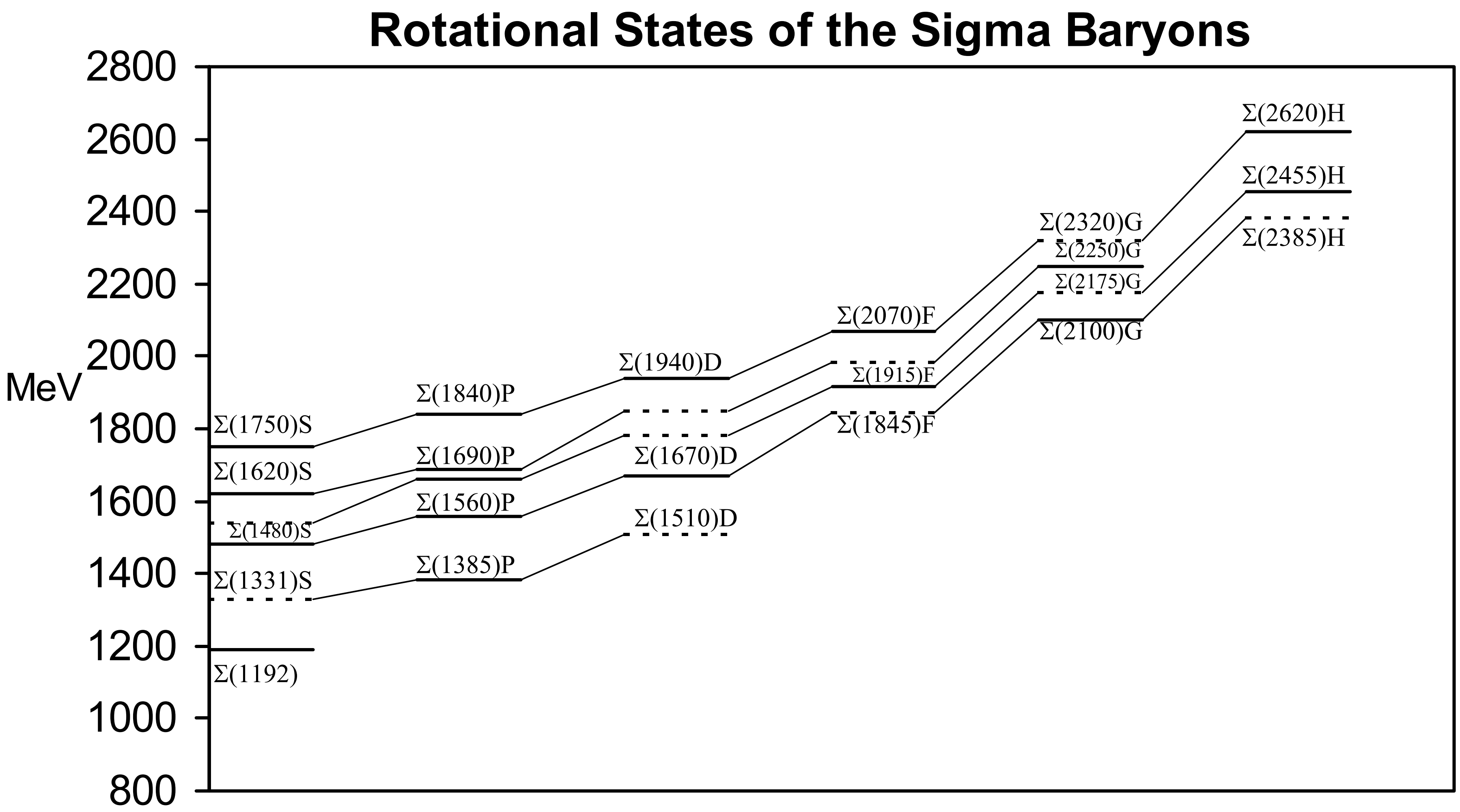

Fig. 6

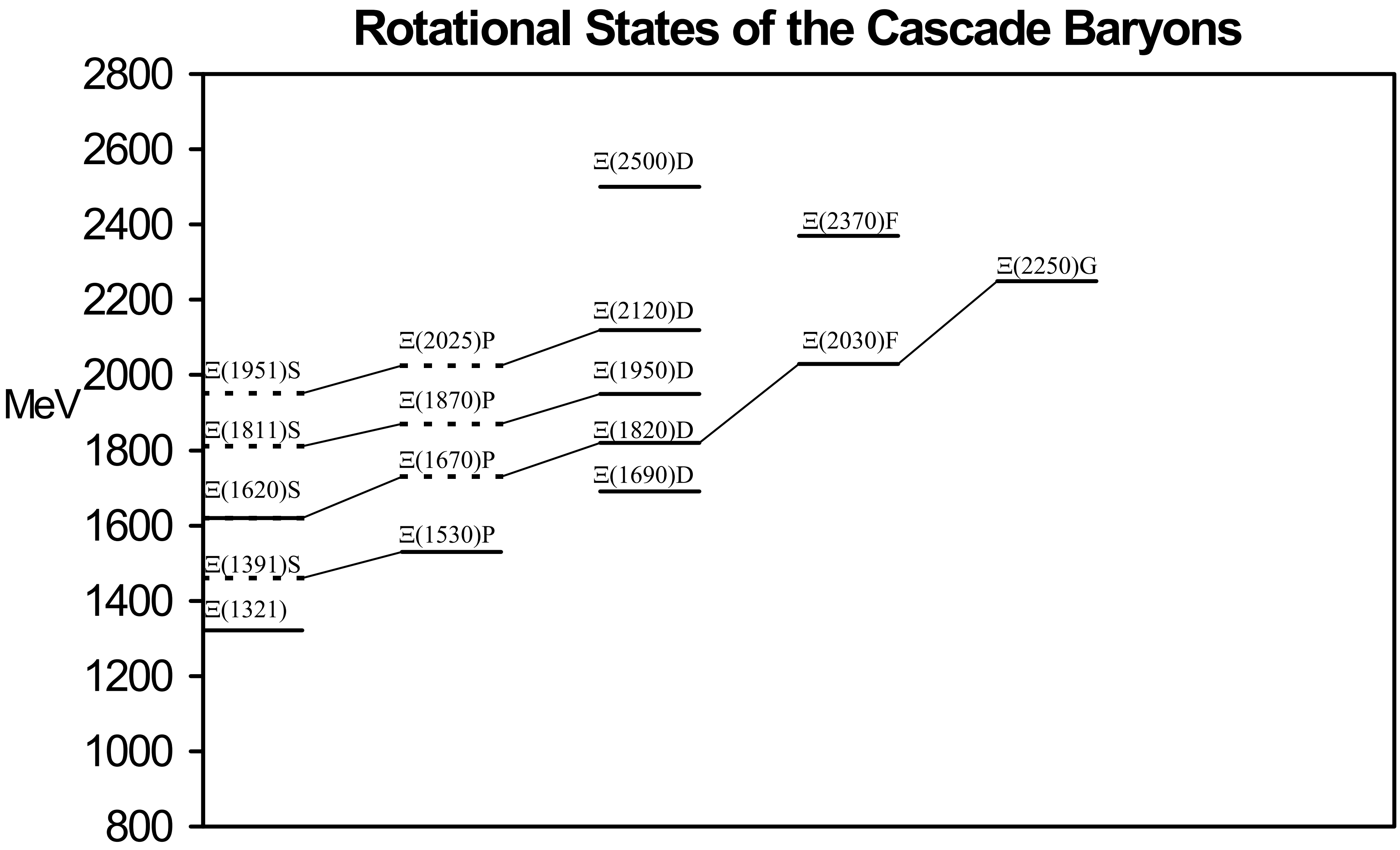

Fig. 7

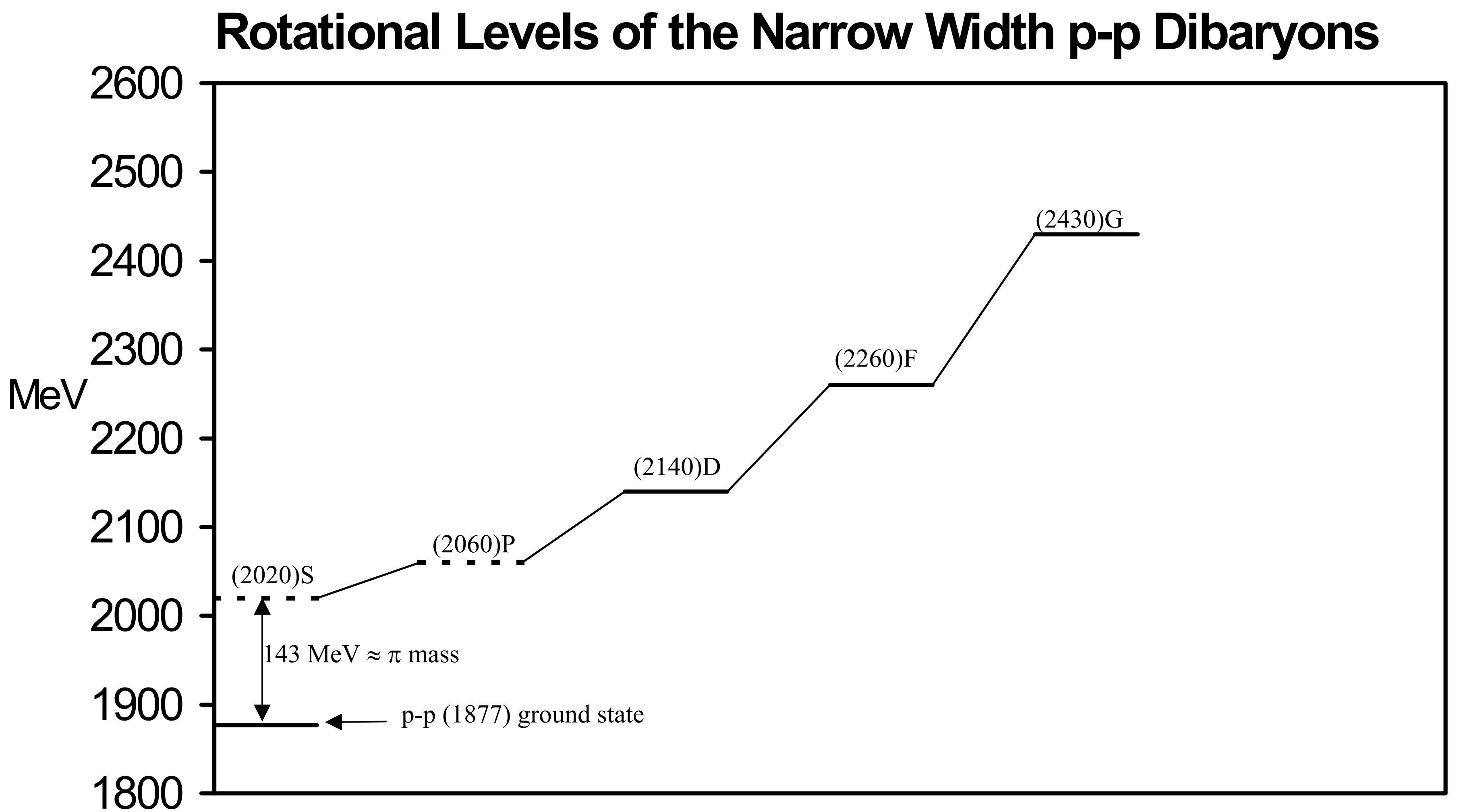
Rotational Levels of the Narrow Width p-p Dibaryons
2600
2500
2400
2300
2200
2100
2000
1900
1800
MeV
(2430)G
(2260)F
(2140)D
(2060)P
(2020)S
143 MeV ≈ π mass
p-p (1877) ground state

Fig. 8

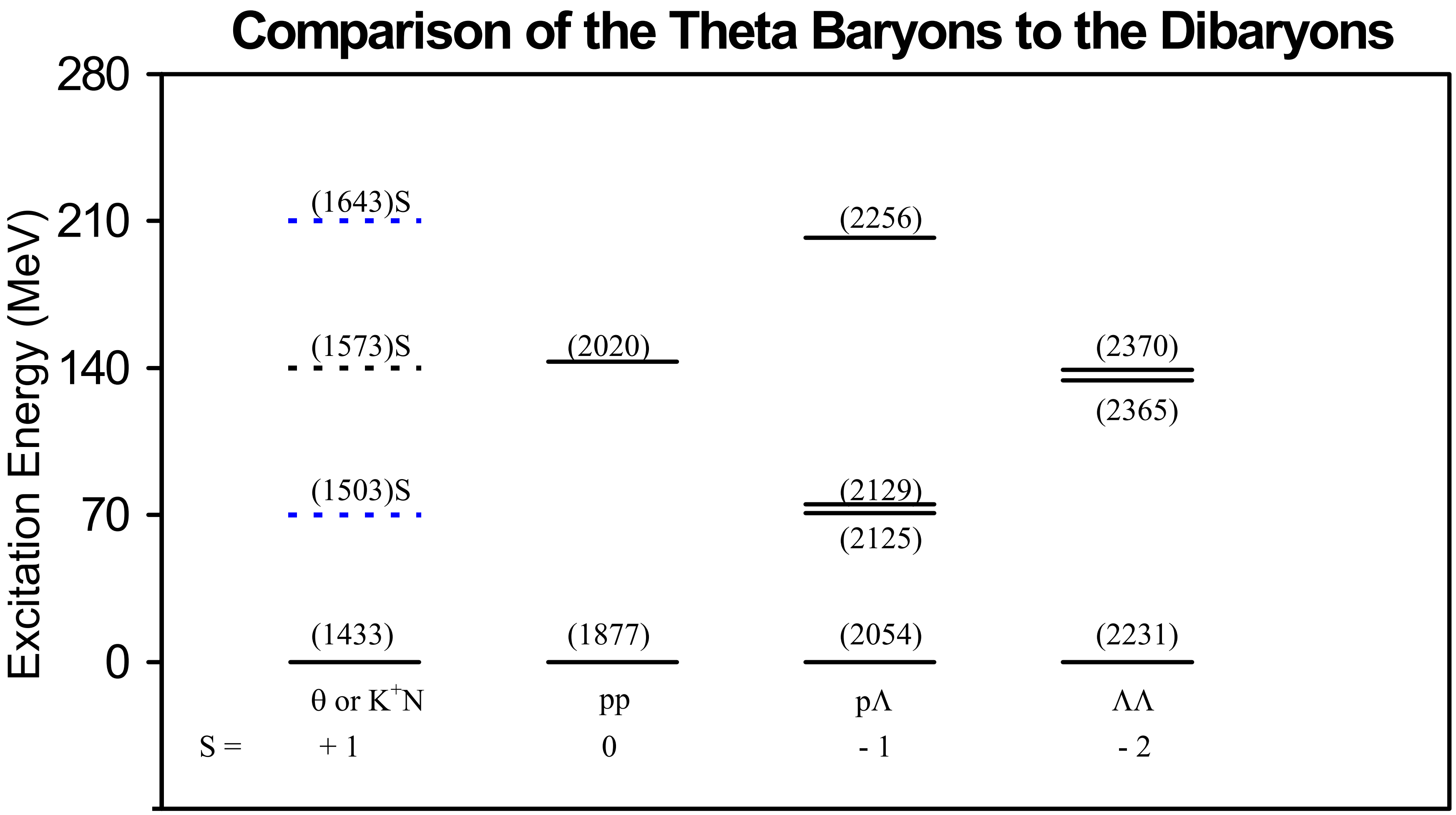
Comparison of the Theta Baryons to the Dibaryons
Excitation Energy (MeV)
280
210
140
70
0
(1643)S
(1573)S
(1503)S
(1433)
(2020)
(1877)
(2256)
(2129)
(2125)
(2054)
(2370)
(2365)
(2231)
θ or K+N
pp
pΛ
ΛΛ
S =
+ 1
0
- 1
- 2

Fig. 9

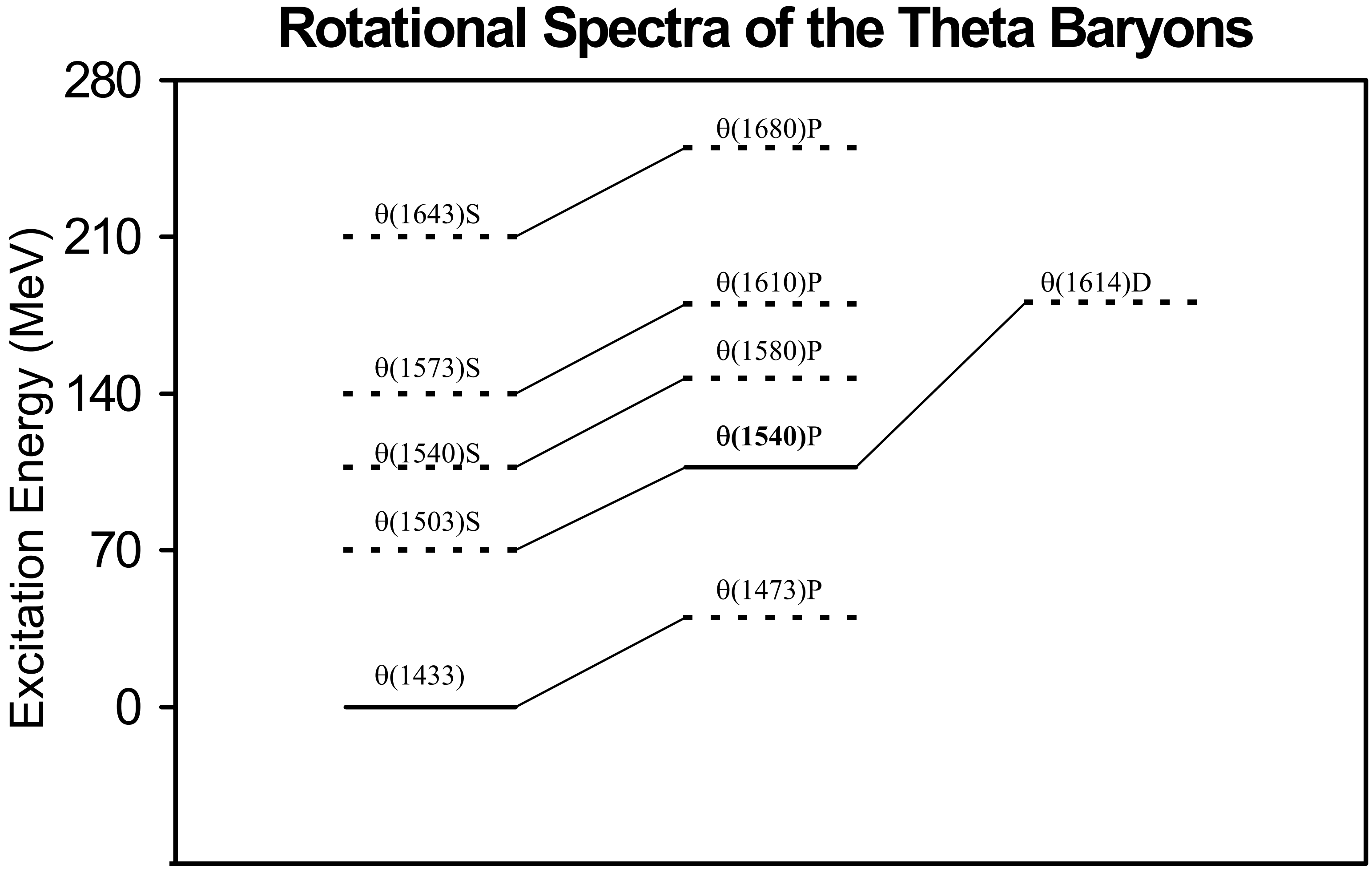

Fig. 10

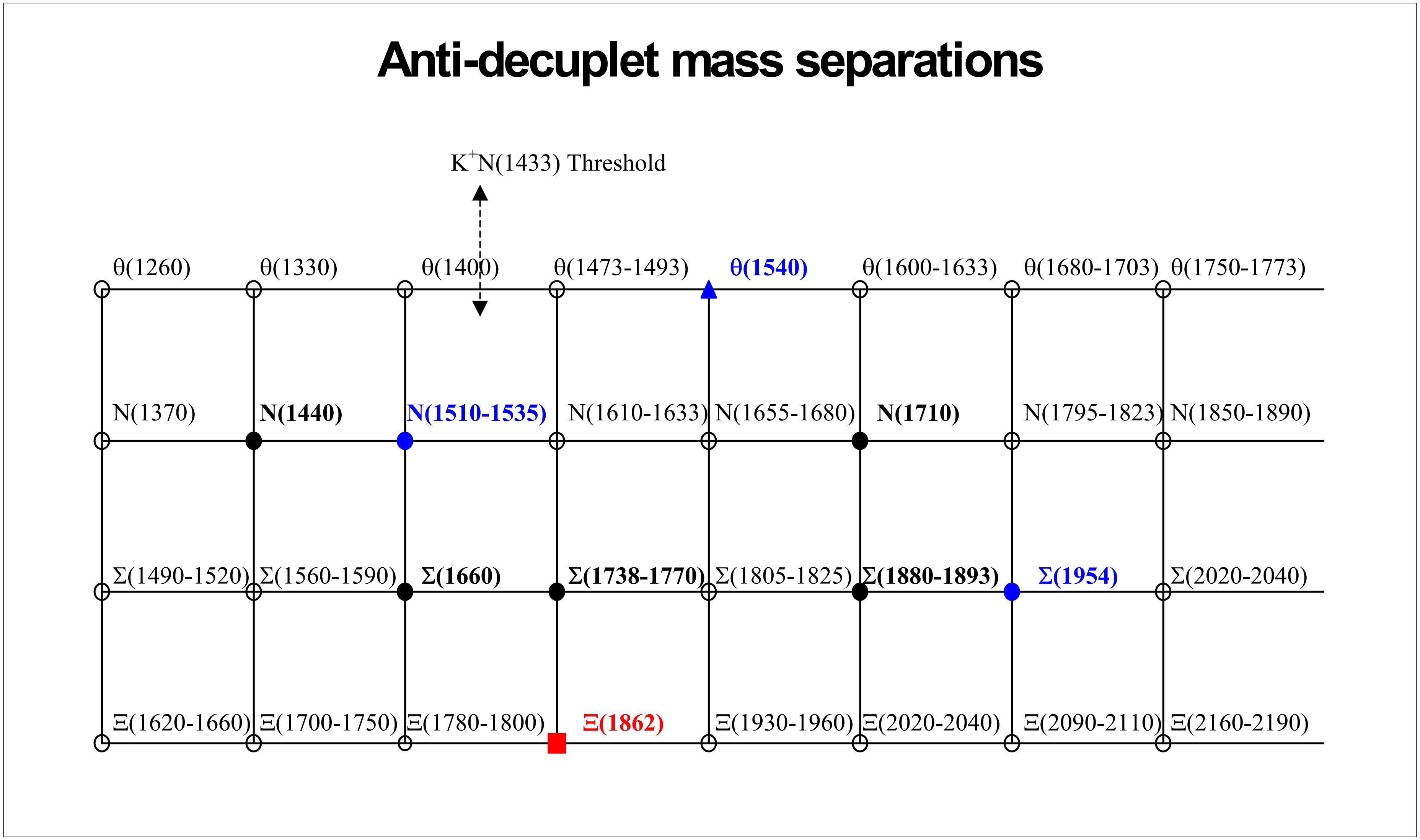

Fig. 11

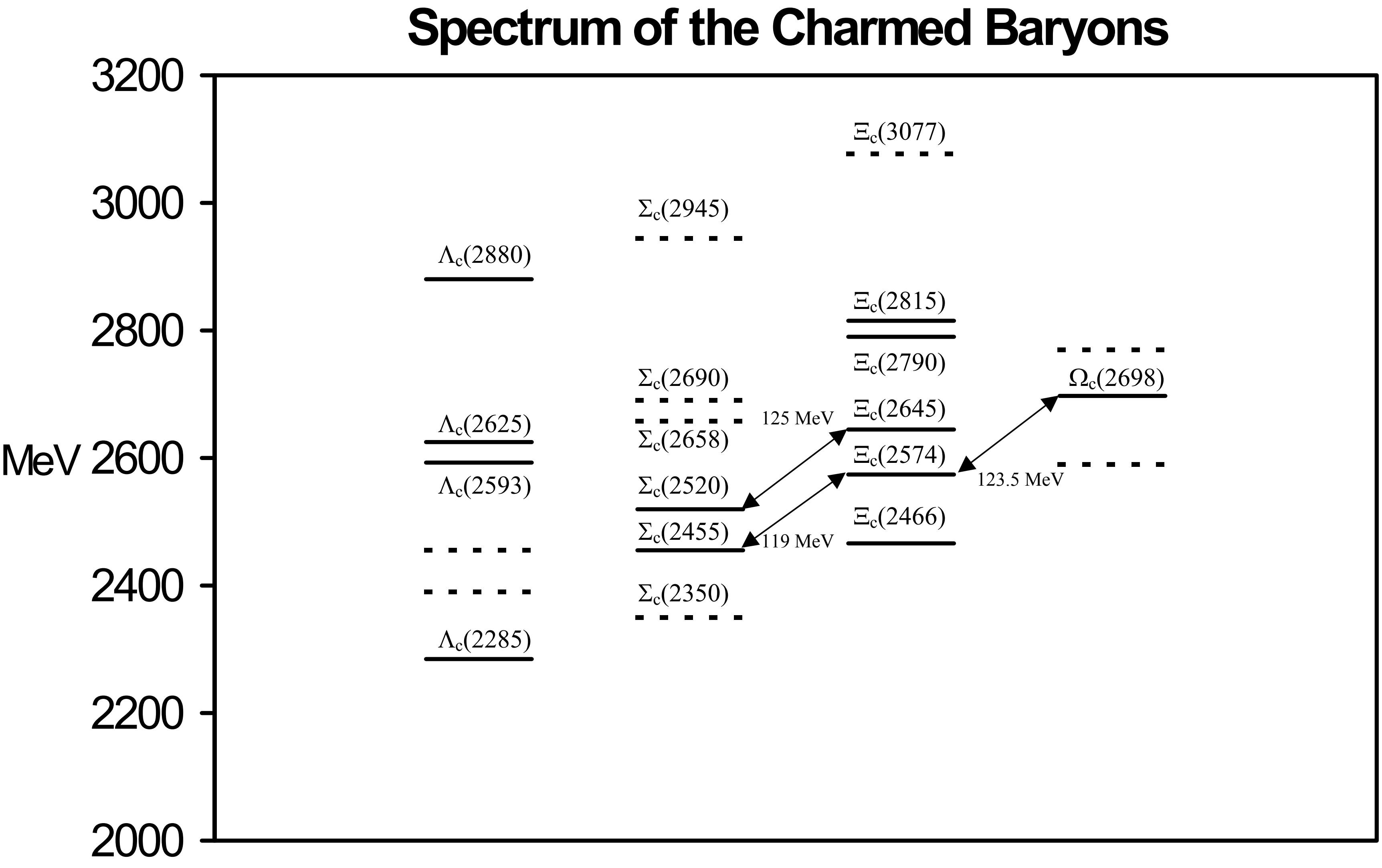

Fig. 12